\documentclass[prl,twocolumn]{revtex4}
\usepackage{bm}

\def\OO         {{\cal O}}
\def\half	{\textstyle {1\over2}}
\def\Eq#1{Eq.~(\ref{#1})}
\def\nsp        {\hspace{-0.5cm}}

\begin{document}
\title{Kubo Formulae for Second-Order Hydrodynamic Coefficients}
\author{Guy D.~Moore and Kiyoumars A.~Sohrabi}
\affiliation
    {%
    Department of Physics,
    McGill University, 3600 rue University,
    Montr\'eal QC H3A 2T8, Canada
    }%
\date{\today}

\begin{abstract}
At second order in gradients, conformal relativistic hydrodynamics
depends on the viscosity $\eta$ and on five additional ``second-order''
hydrodynamical coefficients
$\tau_\Pi$, $\kappa$, $\lambda_1$, $\lambda_2$, and $\lambda_3$.
We derive Kubo relations for these coefficients, relating them to
equilibrium, fully retarded 3-point correlation functions of the stress
tensor.  We show that the coefficient $\lambda_3$ can be evaluated
directly by Euclidean means and does {\sl not} in general vanish.
\end{abstract}


\maketitle


\vspace*{-5pt}

Results from the RHIC experiments, particularly the
measurement of a large transverse flow \cite{experiments}, appear to
show that the Quark-Gluon plasma can be well described by hydrodynamics
with a surprisingly small viscosity \cite{ideal_hydro}.  A major future
goal for heavy ion experiment and theory is to quantify how small the
viscosity of the plasma is.  This requires the numerical treatment of
relativistic viscous hydrodynamics \cite{visc_hydro}.  It has long been
known \cite{Muller,IsraelStewart,Hiscock} that the relativistic Navier-Stokes
equations are acausal and unstable.  But the Navier-Stokes equations are
just the result of a first-order (Chapman-Enskog \cite{vanweert})
expansion in gradients.  Extending the expansion to second order yields
numerically stable equations after a certain reorganization is applied
\cite{IsraelStewart}.  The drawback is that
it adds unknown coefficients.  In the conformal case (which we will
consider for simplicity), besides the equation of state $P(\epsilon)$ at
zero order and the shear viscosity $\eta$ at first order, there are five new
transport coefficients: $\tau_\Pi,\kappa,\lambda_1,\lambda_2, \lambda_3$
in the notation of \cite{BRSSS}.  These have been evaluated in strongly
coupled ${\cal N}{=}4$ super-Yang-Mills theory in the limit of many colors
\cite{BRSSS,Tata} and at leading order in weakly coupled QCD
\cite{MooreYork}. In each case $\lambda_3 = 0$ at lowest order in the
respective (strong or weak coupling) expansion.

Baier {\it et al} have also presented Kubo formulae for two of these
coefficients, $\tau_\Pi$ and $\kappa$, which relate them to well
defined, equilibrium correlation functions of the stress tensor.
Presumably, the remaining three coefficients $\lambda_{1,2,3}$ can also
be expressed in terms of stress tensor correlation functions.  Doing so
would put the definition of these coefficients on a solid footing and
might aid in their physical interpretation and their theoretical
calculation.  In the remainder of this paper we will derive such Kubo
relations for the three remaining second-order coefficients.
We do this first by showing how the first and second order hydrodynamic
coefficients can be related to the stress tensor in a background
spacetime with perturbatively small geometrical curvature.  Then we
expand in the metric as an external background field {\it a la} Kubo
\cite{Kubo} and derive a relation between $\lambda_{1,2,3}$ and certain
fully retarded 3-point stress-tensor correlation functions.  This allows
us to determine the previously unknown perturbative behavior of the
coefficient $\lambda_3$ (which is not zero) and to say something about
its physical interpretation.

We restrict attention to conformal fluids mostly to simplify the
presentation; in the nonconformal case there are more coefficients
\cite{Romatschke} but there are no conceptual or technical obstacles to
treating this case with the same methodology developed here.

\section{Constitutive Relations for Second Order Coefficients}

We begin by {\sl defining} the second order coefficients.
The expectation value of the
stress-energy tensor operator for a fluid can be decomposed in terms
of a local equilibrium piece and an extra piece,
\begin{eqnarray}
\label{Tmunu}
\langle T^{\mu\nu} \rangle & = & T^{\mu\nu}_{\rm eq}(u^\mu,\epsilon)
 +\Pi^{\mu\nu} \,, \nonumber \\
T^{\mu\nu}_{\rm eq} & \equiv & (\epsilon+P) u^\mu u^\nu + P g^{\mu\nu}\,.
\end{eqnarray}
Here $g_{\mu\nu},\epsilon,P,u^\mu$ are the spacetime metric (in the
mostly-plus convention), energy density, pressure as given by the
equation of state, and flow
4-velocity.  We work in the Landau-Lifshitz frame,
$u_\mu \Pi^{\mu\nu} = 0$, which makes the division between
$T^{\mu\nu}_{\rm eq}$ and $\Pi^{\mu\nu}$ unique; we normalize
$u^\mu$ so that $u_\mu u^\mu = -1$.
 While $u^\mu$, $\epsilon$, and $g_{\mu\nu}$ are ordinary functions of
 $x$, $T^{\mu\nu}$ is a Heisenberg-picture operator;
 $\langle T^{\mu\nu} \rangle$ represents its trace in the density matrix
 describing the fluid.

The key idea of hydrodynamics is that, for a system which varies slowly
in space and time, $\Pi^{\mu\nu}$ arises only due to the
nonuniformity of the system and should therefore be expressible in terms
of a gradient expansion in that nonuniformity.
To write out $\Pi^{\mu\nu}$ to second order, we introduce some notation.
We define $\Delta^{\mu\nu} \equiv g^{\mu\nu} + u^\mu u^\nu$, which is
the projector to spatial directions in the local rest frame.
Angular brackets around a pair of Lorentz indices,
${}^{\langle \mu\nu \rangle}$, mean that the indices are to be
symmetrized, space-projected, and trace-subtracted; that is,
\begin{equation}
A^{\langle \mu \nu \rangle} \equiv
\frac{1}{2} \Delta^{\mu\alpha} \Delta^{\nu\beta}
( A_{\alpha\beta} + A_{\beta\alpha} )
- \frac{1}{3} \Delta^{\mu\nu} \Delta^{\alpha\beta} A_{\alpha\beta} \,.
\end{equation}
The shear and vorticity tensors are defined as
\begin{eqnarray}
\sigma^{\mu\nu} & \equiv & 2 \nabla^{\langle \mu } u^{\nu \rangle} \,,
\\
\Omega^{\mu\nu} & \equiv & \half \Delta^{\mu\alpha} \Delta^{\nu\beta}
( \nabla_\alpha u_\beta - \nabla_\beta u_\alpha) \,.
\end{eqnarray}
$R^{\mu\nu}$ and $R^{\mu\nu\alpha\beta}$ are the Ricci tensor and
curvature tensor respectively.  In terms of these quantities,
the most general form for $\Pi^{\mu\nu}$ compatible with conformal
symmetry is \cite{BRSSS}
\begin{eqnarray}
  \label{viscosity}
\Pi^{\mu\nu} & = & - \eta\sigma^{\mu\nu}
   + \eta \tau_{\Pi}\left(u\cdot \nabla
                  \sigma^{\langle \mu\nu \rangle}
                + \frac{\nabla\cdot u}{3}\sigma^{\mu\nu}\right)
\nonumber \\ &&
 + \kappa\left( R^{\langle\mu\nu\rangle}
     - 2u_{\alpha}u_{\beta}R^{\alpha\langle\mu\nu\rangle\beta}\right)
\\ &&
+ \lambda_{1}\sigma_{\lambda}{}^{\langle\mu} \sigma^{\nu\rangle\lambda}
+\lambda_{2}\sigma_{\lambda}{}^{\langle\mu}\Omega^{\nu\rangle\lambda}
-\lambda_{3}\Omega_{\lambda}{}^{\langle\mu}\Omega^{\nu\rangle\lambda}
  \,. \nonumber
\end{eqnarray}

\section{Expansion in Background geometry}

We derive Kubo relations for $\lambda_1$ {\it etc.}\ by considering a
system where some nonuniformity, either in the initial conditions or in
the spacetime geometry, forces $\sigma^{\mu\nu}$ {\it etc.}\ to be
nonzero.  It is particularly convenient to consider an initially
uniform, equilibrium system in flat space but to introduce
perturbatively weak and slowly varying
spacetime nonuniformity which causes the fluid to experience shear
and vorticity.  Writing the metric as
$g_{\mu\nu}(x) = \eta_{\mu\nu} + h_{\mu\nu}(x)$ ($\eta_{\mu\nu}$ the
flat-space metric), one expands perturbatively in $h_{\mu\nu}$.
Since $h_{\mu\nu}$ couples to the
stress tensor $T^{\mu\nu}$, this generates an expansion in correlation
functions of multiple stress tensors, whose coefficients are the response
of the stress tensor to fluid nonuniformities.

Consider the expectation value
$\langle T^{\mu\nu}(0) \rangle$ for a system initially (time $t_0\ll 0$) in
equilibrium at temperature $T$, subject to a spacetime dependent metric
perturbation $h_{\alpha\beta}(x)$, with $h_{\mu\nu}(t\leq t_0)=0$.  The
stress tensor is determined by
\begin{eqnarray}
\langle T^{\mu\nu}(0) \rangle & = & {\rm Tr}\: e^{-\beta H}
 {\rm \tilde{T}exp}\left( \int_{t_0}^0  dt' i H[h(t')] \right)
T^{\mu\nu} \nonumber \\ && \times
{\rm Texp}\left(  \int_{t_0}^0 dt'' (-i) H[h(t'')] \right)
\end{eqnarray}
(with ${\rm \tilde{T}exp}$ and ${\rm Texp}$ the anti-time ordered and
time-ordered exponentials respectively, $H[h(t)]$ the Hamiltonian,
showing explicitly its dependence on the metric,  and $\beta{=}T^{-1}$
the inverse temperature).
This is best treated using the Schwinger-Keldysh (closed
time path) formalism  (see \cite{ChouSuHaoYu,WangHeinz};
we follow the conventions in \cite{WangHeinz}).
We introduce {\sl independent} metric perturbations for the
${\rm T}$-ordered and $\tilde{\rm T}$-ordered evolution operators in the
above expression and define the generating functional
\begin{eqnarray} \!\!\!\!\!
W[h_1,h_2] & \equiv & \ln {\rm Tr}\: e^{-\beta H}
{\rm \tilde{T}exp} \left(i \int_{t_0}^\infty dt' H[h_2(t')] \right)
\nonumber \\ && \qquad \times
{\rm Texp} \left( -i \int_{t_0}^\infty dt' H[h_1(t')] \right)
 \\ & = &
\ln \int {\cal D}[\Phi_1,\Phi_2,\Phi_3]
e^{i\int \sqrt{-g_1} d^4 x {\cal L}[\Phi_1(x),h_1]}
\nonumber \\ && \times
e^{-\int_0^\beta \!\!\! d^4 z {\cal L}_{\rm E} [\Phi_3(z)]}
e^{-i\int \sqrt{-g_2} d^4 y {\cal L}[\Phi_2(y),h_2]} \,. \nonumber
\end{eqnarray}
One then defines the average metric perturbation
$h_r \equiv \frac{h_1+h_2}{2}$ and stress tensor
$T_r \equiv \frac{T_1+T_2}{2}$, and the difference variables
$h_a \equiv h_1-h_2$, $T_a \equiv T_1-T_2$.  Variation with respect
to $h_a$ gives $T_r$, explicitly
\begin{equation}
\frac{-2i}{\sqrt{-g}} \frac{\partial W}{\partial h_{a\mu\nu}(x)}
= \langle T^{\mu\nu}_r(x) \rangle \,.
\end{equation}
We use such a variation to pull down the $T^{\mu\nu}(0)$ factor we want
to evaluate.  After taking this $h_a$ derivative, we set $h_a=0$
and $h_r = h$, since we are interested in the case of a classical
background value $h_1=h_2=h_r=h$.
(The difference $h_a$ represents possible
quantum fluctuations in the metric which we do not want to consider.)
We then expand order by order in $h_{r\mu\nu}$ to obtain a series
expansion of $\langle T_{r}^{\mu\nu} \rangle$ in powers of $h$.
Explicitly, we find
\begin{eqnarray}
\label{Texpand}
\langle T_r^{\mu\nu}\rangle_{h} & = & G_r^{\mu\nu}(0)
- \frac{1}{2} \int d^4 x G_{ra}^{\mu\nu,\alpha\beta}(0,x) h_{\alpha\beta}(x)
\\ && \nonumber
 \! + \frac{1}{8} \int d^4 x d^4 y \,
 G_{raa}^{\mu\nu,\alpha\beta,\gamma\delta}(0,x,y)
        h_{\alpha\beta}(x) h_{\gamma\delta}(y) \;
\end{eqnarray}
plus terms of order $h^3$.
Here $G^{\mu\nu,\alpha\beta,\ldots}_{ra \ldots}(0,x,\ldots)$ is the
correlation function of one $T_r$ and 0 or more $T_a$'s,
\begin{eqnarray} \hspace{-0.2em}
G^{\mu\nu,\alpha\beta,\ldots}_{r a \ldots}(0,x,\ldots) & \!\equiv\!\! &
\left. \frac{(-i)^{n-1}(-2i)^n \partial^n W}{\partial g_{a,\mu\nu}(0)
         \partial g_{r,\alpha\beta}(x) \ldots }
        \right|_{g_{\mu\nu}=\eta_{\mu\nu}}
\\ & \!=\!\! & (-i)^{n-1}
\left\langle T^{\mu\nu}_r(0) T^{\alpha\beta}_a(x) \ldots
     \right\rangle_{\rm eq} \!
+\mbox{c.t.}  \nonumber
\end{eqnarray}
The expectation value is with respect to the flat-space, equilibrium
density matrix.  $G_{ra\ldots}$ is a fully retarded correlation
function \cite{WangHeinz},
which is a nested commutator, from earliest to latest time,
with $T_r$ at the last time and innermost in the commutator,
{\it eg} when $x^0<y^0<\ldots<0$ the correlator is
$\langle [T(x),[T(y),[\ldots T(0)]]]\rangle$.
Here (c.t.) refers to the contact terms which are built into our
definition of the n-point stress tensor correlation functions.
This is discussed in \cite{RomatschkeSon}; the contact terms turn out
not to be important for evaluating $\eta$ but they will contribute to
the evaluation of $\lambda_{1,2,3}$.

\section{Kubo formulae}

First we review the derivation of Kubo formulae for the ``linear''
transport coefficients $\eta,\tau_\Pi,\kappa$ \cite{BRSSS}.
Consider $\langle T^{xy}\rangle$ in the presence of
$h_{xy}(z,t)$.  According to \Eq{Texpand}, at first order
\begin{equation}
\label{G2_1}
\langle T_r^{xy} \rangle_{h} = - \int d^4 x \; h_{xy}(x)
G^{xy,xy}_{ra}(0,x) + {\cal O}(h^2) \,.
\end{equation}
Using \Eq{Tmunu}\ and $\nabla_\mu T^{\mu\nu}=0$ (energy-momentum
conservation), we derive that $u^i=0$ at
${\cal  O}(h)$.  We then evaluate $\sigma^{xy}$,
$u\cdot \nabla \sigma^{\langle \mu\nu \rangle}$ {\it etc.}\
explicitly for this $u^{\mu}$ and $h_{\mu\nu}$,
finding for instance that
$\sigma^{xy} = \partial_t h_{xy}$.  Substituting into
\Eq{viscosity}, we find
\begin{eqnarray}
\langle T_r^{xy} \rangle_h &=& -P h_{xy} -\eta \partial_t h_{xy}
  + \eta \tau_\Pi \partial_t^2 h_{xy}
\nonumber \\ &&
-\frac{\kappa}{2} \left( \partial_z^2 h_{xy} + \partial_t^2 h_{xy}
\right) + {\cal O}(\partial^3,h^2) \,.
\label{G2_2}
\end{eqnarray}
defining $G_{ra}^{xy,xy}(\omega,k)= \int d^4 x e^{i(\omega t-kz)}
G^{xy,xy}_{ra}(0,-x)$
and equating Eqs.~(\ref{G2_1},\ref{G2_2}) order by order in derivatives,
we find
\begin{eqnarray}
\label{Kuboeta}
\hspace{-1em}
\eta & \!=\! & i \partial_\omega G^{xy,xy}_{ra}(\omega,k)|_{\omega=0=k}
\,, \\  \hspace{-1em}
\label{Kubokappa}
\kappa & \!=\! & -\partial^2_{k_z} G^{xy,xy}_{ra}(\omega,k)|_{\omega=0=k}
\,, \\  \hspace{-1.6em}
\label{Kubotau}
\eta \tau_\pi & \!=\! & \frac{1}{2} \! \left.
   \left( \partial^2_{\omega}  G^{xy,xy}_{ra}(\omega,k)
         -\partial^2_{k_z} G^{xy,xy}_{ra}(\omega,k)
    \right) \right|_{\omega=0=k}.
\end{eqnarray}
These reproduce the Kubo relations obtained by \cite{BRSSS}.

To obtain higher order Kubo formulae for the nonlinear coefficients, we
continue this procedure to $\OO(h^2)$, for a background
choice which allows nonzero shear flow and vorticity.  To do so, we will
consider $\Pi^{xy}$ arising when $h_{xz}(t)$, $h_{yz}(t)$, $h_{x0}(z)$,
and $h_{y0}(z)$ are nonvanishing%
\footnote%
{%
   Technically $h_{x0},h_{y0}$ must depend on $z,t$ since it must vanish in the
   initial conditions.  It is essential to ``turn on'' this perturbation
   very slowly, on a timescale $t \gg (\epsilon+P)/(k^2 \eta)$
   ($k$ the wave number for $h_{x0}$),
   and to include the viscous term $-\eta \sigma^{\mu\nu}$ in
   \Eq{viscosity}, to correctly derive that $u^i=0$ after fully turning
   on the $h_{x0},h_{y0}$ perturbations.
}.
By our choice of indices, $\sigma_{xz}=\partial_t h_{xz}$,
$\sigma_{yz}=\partial_t h_{yz}$, $\Omega_{xz}=-\partial_z h_{x0}/2$
and $\Omega_{yz}=-\partial_z h_{y0}/2$ arise at $\OO(h)$, but
$\Pi^{xy}$ will automatically only arise at order in $h^2$, and will not
receive contributions at this order from the $(\epsilon{+}P) u^x u^y$
term.  Since $g^{xy}$ is explicitly $\OO(h^2)$, we also only need the
equilibrium value of $P$%
\footnote%
{%
    We are indebted to Peter Arnold, Diana Vaman, Chaolun Wu, and Wei
    Xiao for pointing out an error in the original version of this
    paper (see \cite{Vaman}).
    At this point we proposed investigating $\Pi^{xx}$ using
    nontrivial $h_{xy}(z,t)$, $h_{x0}(y)$.  However, in this case
    $g^{xx}$ arises at $\OO(h^0)$, and so $\OO(h^2)$ corrections to the
    pressure $P$ must be evaluated.  We failed to do so, and therefore
    the Kubo relations in the original version of this paper were in
    error.
}.
Explicitly
evaluating \Eq{viscosity} in this background to second order, we find
\begin{eqnarray}
\label{pixx}
\label{stress-xx}
\!\!\langle T^{xy}\rangle & \!=\! &
   P \left( h_{xz} h_{yz}- h_{x0} h_{y0} \right)
  +\eta \left( h_{xz} h_{yz,t} + h_{xz,t} h_{yz} \right)
\nonumber \\ &+&
  \frac{\kappa}{2} \left( h_{xz} h_{yz,tt} {+} h_{xz,tt} h_{yz}
                           {-}h_{xt} h_{yt,zz} {-} h_{xt,zz} h_{yt} \right)
\nonumber \\ &+&
  \frac{\eta \tau_\pi}{2} 
   \left( h_{xz,t} h_{yt,z} + h_{yz,t} h_{xt,z} - 2h_{xz,t} h_{yz,t}
\right. \nonumber \\ && \left. \quad
         -2h_{yz} h_{xz,tt} - 2h_{xz} h_{yz,tt} \right)
\nonumber \\ &+&
  \lambda_1 \left( h_{xz,t} h_{yz,t} \right)
  -\frac{\lambda_2}{4} \left( h_{xz,t} h_{yt,z} 
                            + h_{yz,t} h_{xt,z} \right)
\nonumber \\ &+&
  \frac{\lambda_3}{4} \left( h_{xt,z} h_{yt,z} \right) \,.
\end{eqnarray}
Equating with the $h^2$ part of \Eq{Texpand}, and defining
\begin{equation}
G^{\mu\nu,\alpha\beta,\sigma\lambda}_{raa}(p,q)
\equiv \int d^4 x d^4 y e^{-i(p\cdot x + q\cdot y)}
G^{\mu\nu,\alpha\beta,\sigma\lambda}_{raa}(0,x,y)
\end{equation}
we find the following Kubo relations:
\begin{eqnarray}
\label{lambda1}
\nsp \lambda_1 &\!=\!& \eta \tau_\pi - \lim_{p^0,q^0 \rightarrow 0}
\partial_{p^0} \partial_{q^0} \lim_{\vec p,\vec q\rightarrow 0}
G^{xy,xz,yz}_{raa}(p,q) \,, \\
\label{lambda2}
\nsp \lambda_2 &\!=\!& 2\eta \tau_\pi {-} 4 \lim_{p^0,q^z \rightarrow 0}
\partial_{p^0} \partial_{q^z} \lim_{\vec p,q^{0,x,y}\rightarrow 0}
G^{xy,xz,yt}_{raa}(p,q) , \\
\label{lambda3}
\nsp \lambda_3 &\!=\!& -4 \lim_{p^z,q^z \rightarrow 0}
\partial_{p^z} \partial_{q^z} \lim_{p^{0,x,y},q^{0,x,y}\rightarrow 0}
G^{xy,xt,yt}_{raa}(p,q) \,.
\end{eqnarray}
These Kubo relations are our main result.

We also find {\sl extra} Kubo relations for $\eta$, $\kappa$, and
$\tau_\pi$:
\begin{eqnarray}
\label{Kuboeta2}
i \eta & = & \lim_{p^0 \rightarrow 0}
    \frac{\partial}{\partial p^0}
G^{xy,xz,yz}_{raa}(p,q) \,,
\\
\label{Kubokappa2}
\kappa & = & 2 \lim_{p^z \rightarrow 0}
    \frac{\partial^2}{(\partial p^z)^2}
G^{xy,x0,y0}_{raa}(p,q) \,,
\\
\label{Kubotau2}
2\eta\tau_\pi - \kappa & = & 2 \lim_{p^0 \rightarrow 0}
    \frac{\partial^2}{(\partial p^0)^2}
G^{xy,xz,yz}_{raa}(p,q) \,,
\end{eqnarray}
where besides the differentiated variable all other $p,q$ components are
taken to zero first.
These extra relations require inter-relations between
$G^{\mu\nu,\alpha\beta}_{ra}$ and
$G^{\mu\nu,\alpha\beta,\gamma\delta}_{raa}$.
Each extra Kubo formula involves one stress tensor at {\sl zero}
external 4-momentum, arising from an undifferentiated $h_{\mu\nu}$
in \Eq{pixx}.  We can always force $h_{\mu\nu}=0$ at $x=0$ where
$T^{xy}$ is evaluated by a coordinate ``gauge'' choice.  The invariance
of the theory to such gauge choice enforces (Ward) relations between two
point functions and three point functions with a $T^{\mu\nu}_a$ at zero
4-momentum.  Consider a stress tensor two-point function
in a spacetime-independent, background $h_{\mu\nu}$:
\begin{eqnarray}
\label{eq:gauge1}
\langle T^{\mu\nu}_r(0) T^{\alpha\beta}_a(x) \rangle_h
& \!=\! &
iG^{\mu\nu,\alpha\beta}_{ra}(0,x)
\\ &&
- \frac{i}{2} \!\int\! d^4 y \; h_{\gamma\delta}(y)
G^{\mu\nu,\alpha\beta,\gamma\delta}_{raa}(0,x,y) \,. \nonumber
\end{eqnarray}
The gauge change which eliminates $h_{\mu\nu}$ is
$x^\mu \rightarrow x^{\mu} + \xi^\mu$ with
$\xi_{\mu,\nu}+\xi_{\nu,\mu} = h_{\mu\nu}$.  Applying the gauge change
to the lefthand side of \Eq{eq:gauge1}, we re-express it in terms of
$h_{\mu\nu}$ and the flat-space correlation functions;
\begin{eqnarray}
&&
h_{\gamma\delta} \Big(
\left[ \eta^{\mu\gamma}G^{\delta\nu,\alpha\beta}_{ra}(p)
  + (\mu\leftrightarrow \nu) \right]
\nonumber \\ && \;\;
+ \left[ \eta^{\alpha\gamma}G^{\mu\nu,\delta\beta}_{ra}(p)
+ (\alpha\leftrightarrow \beta) \right]
+(\gamma\leftrightarrow \delta) \Big)
\nonumber \\
& = &
2 h_{\gamma\delta} G^{\mu\nu,\alpha\beta,\gamma\delta}_{raa}
(p,0) \,.
\end{eqnarray}
Choosing $\mu\nu=xy$, $\alpha\beta = xz$, $\gamma\delta=yz$,
\begin{equation}
G^{xy,xy}_{ra}(p) + G^{xz,xz}_{ra}(p)
=2 G^{xy,xz,yz}_{raa}(p,0) \,.
\label{above}
\end{equation}
Now $\partial_\omega G^{xy,xy}=i\eta$ by \Eq{Kuboeta}
and $\partial_\omega G^{xz,xz}=i\eta$ by rotational invariance,
so\Eq{Kuboeta2} follows.  The same procedure applies for the other
linear coefficients.

\section{Discussion}

Our derivation had two goals.  First, we wanted relations,
shown in Eqs.~(\ref{lambda1},\ref{lambda2},\ref{lambda3}), for the
second-order nonlinear transport coefficients in terms of equilibrium
energy-momentum tensors.  Second, we hoped that these relations would
shed some light on the nature or properties of these transport
coefficients.  The most mysterious of these transport coefficients is
$\lambda_3$, which is found to vanish in
${\cal N}{=}4$ SYM theory in the limit of many colors and large
coupling \cite{Tata} and which is zero at order $g^{-8}$ in the weak
coupling expansion, the order where $\lambda_{1,2}$ are nonzero
\cite{MooreYork}.  Is it identically zero?
Romatschke \cite{Romatschke} studied this problem (among others) using a
generalized entropy current and showed that $\lambda_3$ is related to a
certain modification of the entropy density in the presence of
vorticity.  Our Kubo relation allows for a direct evaluation of
$\lambda_3$ in weakly coupled field theory.

$\kappa$ and $\lambda_3$ have expressions involving space
but not time derivatives of stress tensor correlation
functions.  We may immediately set
$\omega=0$ in \Eq{Kubokappa} and $p^0,q^0=0$ in \Eq{lambda3}.  
The frequency-domain fully retarded function $G_{ra\ldots}$ is the
analytic continuation of $G_{\rm E}$ the Euclidean
correlation function.  In particular,
$G_{raa}^{\mu\nu,\alpha\beta,\sigma\tau}(-i\omega_1,-i\omega_2)
= i^{n_0} G_{\rm E}^{\mu\nu,\alpha\beta,\sigma\tau}(\omega_1,\omega_2)$
for Matsubara frequencies
$\omega_{1,2} = 2\pi T n_{1,2}$.  Here $n_0$ is the number of indices
$\mu,\nu,\alpha,\beta,\sigma,\tau$ which are 0, since there is a factor
of $i$ arising from the Euclidean continuation of a 0 index.
This relation shows that the zero-frequency $raa$ and Euclidean
correlation functions are equal up to factors of $i$.  Hence
\begin{eqnarray}
\label{lambda3E}
\lambda_3 = 4 \lim_{\vec p,\vec q \rightarrow 0}
\frac{\partial^2}{\partial p_z \partial q_z} G_{\rm E}^{xy,x0,y0}(p,q)
\,.
\end{eqnarray}
One usually considers such Euclidean correlation functions to carry only
thermodynamical information; $\lambda_3$ should not be thought of as a
dynamical coefficient but as a {\sl thermodynamic} response to
vorticity%
\footnote{For instance, there are curved but time-independent geometries
  and density matrix choices
  where $\kappa$ and $\lambda_3$ contribute to $T^{\mu\nu}$ but the
  fluid is in equilibrium and no entropy production is occurring.}.

At weak coupling we can directly evaluate \Eq{lambda3E} diagrammatically
in the Matsubara formalism.  This contrasts with the case of
$\eta,\tau_\Pi,\lambda_1$, and $\lambda_2$, where time derivatives mean
that $G_{raa}$ must be evaluated at small nonzero frequency where the
continuation cannot be so simply applied.  Therefore the weak coupling
expansion of $\lambda_3$ (and $\kappa$ \cite{RomatschkeSon}) will start at
$g^0$, while the expansions for $\lambda_{1,2}$ can involve inverse
powers of $g$ \cite{Jeon,MooreYork}.

We have evaluated the correlation function in \Eq{lambda3E} for a
one-component scalar field theory at leading order in weak coupling.
Two diagrams contribute; a triangle diagram,
$\partial_{p_z} \partial_{q_z}
   \langle T^{xy}(-p-q) T^{x0}(p) T^{y0}(q)\rangle
   = \frac{T^2}{144}$
and a contact term involving 
$X^{x0y0}\equiv 2\partial T^{x0}/\partial g_{y0}$,
$\partial_{p_z} \partial_{q_z}
   \langle T^{xy}(-p-q) X^{x0y0}(p+q) \rangle
   = \frac{T^2}{72}$.  Hence
\begin{equation}
\lambda_3 = \frac{T^2}{12}\,, \quad
\mbox{1 weak-coupled real scalar field.}
\end{equation}
We get the same answer using the scalar field stress
tensor from \cite{Coleman}.
The important observations are that $\lambda_3$ can
be quite easily evaluated at weak coupling via Euclidean techniques, and
the result is not in general zero.

\section*{Acknowledgements}

We 
thank Peter Arnold, Diana Vaman, Chaolun Wu and Wei
Xiao for pointing out an error in the original version of this paper,
and Alessandro Cerioni, Paul Romatschke, Omid Saremi,
and  Dam Son for useful conversations.   This work was supported in part
by the Natural Sciences and Engineering Research Council of Canada.


\end{document}